\documentclass[twocolumn,showpacs,amsmath,pre]{revtex4}

\usepackage{graphicx}
\usepackage{amsmath}
\usepackage{amssymb}
\usepackage{natbib}

\begin{document}

\title{Rods are less fragile than spheres: Structural relaxation in
dense liquids composed of anisotropic particles}

\author{Tianqi Shen$^{1}$}
\author{Carl Schreck$^{1}$}
\author{Bulbul Chakraborty$^2$}
\author{Denise E. Freed$^3$}
\author{Corey S. O'Hern$^{4,1}$}

\affiliation{$^{1}$ Department of Physics, Yale University, New Haven,
  Connecticut 06520-8120, USA}

\affiliation{$^{2}$ Martin Fisher School of Physics, Brandeis
University, Mail Stop 057, Waltham, Massachusetts 02454-9110, USA}

\affiliation{$^{3}$ Schlumberger-Doll Research Center, One Hampshire
Street, Cambridge, Massachusetts 02139, USA}

\affiliation{$^{4}$ Department of Mechanical Engineering \& Materials
  Science, Yale University, New Haven, Connecticut 06520-8286, USA}

\begin{abstract}
We perform extensive molecular dynamics simulations of dense liquids
composed of bidisperse dimer- and ellipse-shaped particles in 2D that
interact via purely-repulsive contact forces. We measure the
structural relaxation times obtained from the long-time $\alpha$-decay
of the self-part of the intermediate scattering function for the
translational and rotational degrees of freedom (DOF) as a function of
packing fraction $\phi$, temperature $T$, and aspect ratio $\alpha$.
We are able to collapse the packing-fraction and temperature-dependent
structural relaxation times for disks, and dimers and ellipses over a
wide range of $\alpha$, onto a universal scaling function ${\cal
F}_{\pm}(|\phi-\phi_0|,T,\alpha)$, which is similar to that employed
in previous studies of dense liquids composed of purely repulsive
spherical particles in 3D. ${\cal F_{\pm}}$ for both the translational
and rotational DOF are characterized by the $\alpha$-dependent scaling
exponents $\mu$ and $\delta$ and packing fraction $\phi_0(\alpha)$
that signals the crossover in the scaling form ${\cal F}_{\pm}$ from
hard-particle dynamics to super-Arrhenius behavior for each aspect
ratio.  We find that the fragility at $\phi_0$, $m(\phi_0)$, decreases
monotonically with increasing aspect ratio for both ellipses and
dimers. For $\alpha > \alpha_p$, where $\alpha_p$ is the location of
the peak in the packing fraction $\phi_J$ at jamming onset, the
rotational DOF are strongly coupled to the translational DOF and the
dynamic scaling exponents and $\phi_0$ are similar for the rotational
and translational DOF.  For $1 < \alpha < \alpha_p$, the translational
DOF become frozen at higher temperatures than the rotational DOF, and
$\phi_0$ for the rotational degrees of freedom increases above
$\phi_J$.  Moreover, the results for the slow dynamics of dense
liquids composed of dimer- and ellipse-shaped particles are
qualitatively the same, despite the fact that zero-temperature static
packings of dimers are isostatic, while static packings of ellipses
are hypostatic.  Thus, zero-temperature contact counting arguments do
not apply to structural relaxation of dense liquids of anisotropic
particles near the glass transition.
\end{abstract}


\maketitle

\section{Introduction}
\label{intro}

One of the hallmarks of glassy behavior in dense liquids is the rapid
increase in the stress and structural relaxation times as the density
is increased or the temperature is lowered toward the glass
transition~\cite{debenedetti}.  Several recent studies have
characterized how the fragility, {\it i.e.} the rate of increase of
the stress and structural relaxation times with decreasing temperature
near the glass transition~\cite{angell}, depends on the packing
fraction in dense liquids of {\it spherical} particles that interact
via purely repulsive contact
potentials~\cite{haxton,haxton2,berthier,berthier2}.  In particular,
Refs.~\cite{berthier,berthier2} have shown that for dense liquids
composed of bidisperse spheres in three dimensions (3D) the
temperature and packing fraction dependence of the structural
relaxation time $\tau$ can be collapsed onto two master curves, one
for $\phi > \phi_0$ and another for $\phi < \phi_0$.  For $\phi <
\phi_0$, the scaling function reduces to hard-sphere behavior $\tau \sim
\exp[A/|\phi_0 -\phi|^{\delta}]$ in the zero-temperature limit
($T\rightarrow 0$), where $A>0$ is a constant and $\delta$ is a
scaling exponent that does not depend on the form of the repulsive
contact potential.  For $\phi > \phi_0$, the temperature dependence of
the structural relaxation time is super-Arrhenius and the fragility
increases with packing fraction. For 3D bidisperse spheres,
$\phi_0\sim 0.635$, which is distinct from the jammed packing fraction
$\phi_J$ for both fast and slow packing-generation
protocols~\cite{berthier,berthier2,liu2,schreck_order}.

Despite the fact that there have been a number studies of the rapidly
growing structural and stress relaxation times near the glass
transition in dense liquids composed of {\it anisotropic}
particles~\cite{letz,pfleiderer,zhang1,kammerer,kammerer1,kammerer2},
there have been few quantitative calculations of the fragility of
dense liquids as a function of the particle shape.  In this article,
we investigate the slow dynamics of both the translational and
rotational degrees of freedom in dense amorphous liquids composed of
bidisperse dimer- and ellipse-shaped particles in two dimensions (2D)
that interact via purely repulsive contact forces as a function of
temperature, packing fraction, and aspect ratio.

We address several important questions: 1) What are the forms for the
structural relaxation times for the translational and rotational
degrees of freedom for dimer- and ellipse-shaped particles as a
function of temperature and packing fraction?  2) Can we identify master
curves for the structural relaxation times for $\phi > \phi_0(\alpha)$
and $\phi < \phi_0(\alpha)$ for each $\alpha$ similar to those found
for spherical particles at $\alpha=1$? 3) Are the slow dynamics of
dense liquids composed of anisotropic, elongated particles more or
less fragile near $\phi_0(\alpha)$ than spherical particles? 4) Do differences
in microscale geometrical features lead to qualitative changes in
structural relaxation in systems composed of anisotropic particles?
To address these questions, we performed extensive molecular dynamics
simulations of supercooled liquids composed of frictionless, purely
repulsive dimer- and ellipse-shaped particles in 2D over a wide range
of temperature $T$, packing fraction $\phi$, and aspect ratio
$\alpha$.

\begin{figure}[h]
\begin{center}
\includegraphics[scale=0.35]{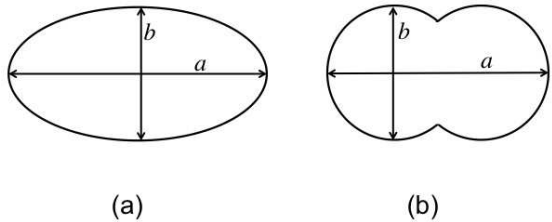}
\vspace{-0.2in}
\caption{Definition of the aspect ratio $\alpha = a/b$, where $a$ and
$b$ are the length of the major and minor axes, for (a) ellipses and
(b) dimers}
\label{axes}
\end{center}
\end{figure}

We find four key results.  First, we show that the structural
relaxation times for dimer- and ellipse-shaped particles in 2D obey
scaling forms similar to that for 3D systems composed of bidisperse
spheres~\cite{berthier,berthier2}.  In particular, we find that there
is an aspect-ratio dependent packing fraction $\phi_0(\alpha)$ below
which the structural relaxation time converges to hard-particle behavior
in the zero-temperature limit and above which the structural
relaxation time grows super-Arrheniusly with decreasing temperature.
Second, we identify a universal scaling form ${\cal
F}_{\pm}(|\phi-\phi_0|,T,\alpha)$ (${\cal F}_-$ for $\phi < \phi_0(\alpha)$
and ${\cal F}_+$ for $\phi > \phi_0(\alpha)$) that collapses the structural
relaxation times for all 2D particle shapes studied, including
bidisperse disks and both dimer- and ellipse-shaped particles over a
wide range of $\alpha$. For small $\alpha$, the rotational degrees of
freedom (DOF) are not `caged' at temperatures where the translational
degrees of freedom become frozen, and $\phi_0(\alpha)$ for the
rotational degrees of freedom increases above $\phi_J(\alpha)$.  For
$\alpha > \alpha_p$ (the location of the peak in
$\phi_J(\alpha)$~\cite{schreck}), the rotational degrees of freedom
are strongly coupled to the translational degrees of freedom, and obey
scaling functions with similar scaling exponents.  Third, the
fragility at $\phi_0$ decreases monotonically with increasing aspect
ratio for both dimers and ellipses. Finally, we do not find
qualitative differences in the slow dynamics of dense liquids composed
of dimer- and ellipse-shaped particles, despite the fact that dimer
packings at zero temperature are isostatic~\cite{witten} at jamming
onset with on average six contacts per particle in 2D ($z=z_{\rm
iso}=6$) for $1 < \alpha \le 2$, whereas ellipse packings at zero
temperature are hypostatic with $z < z_{\rm iso}$ over the same range
in $\alpha$~\cite{schreck}.  Naively, one might have expected that the
structural relaxation times for hypostatic systems decay more quickly
than those for isostatic systems at a given packing fraction
$|\phi-\phi_J(\alpha)|$ and temperature $T$.  However, we show that
counting arguments for zero temperature packings do not apply for dense
liquids composed of anisotropic particles near the glass transition.

\begin{figure}[h]
\begin{center}
\includegraphics[scale=0.35]{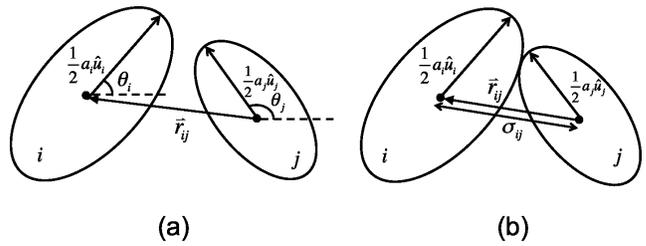}
\caption{(a) Schematic of ellipses $i$ and $j$ with orientations
$\hat{u}_i$ and $\hat{u}_j$, respectively, and interparticle
separation vector $\vec{r}_{ij}$. $\theta_i$ is the angle between
$\hat{u}_i$ and the horizontal axis. (b) Definition of the contact
distance $\sigma_{ij}$ between ellipses $i$ and $j$, which is obtained
by moving ellipse $i$ toward $j$ along ${\hat r}_{ij}$ while fixing
their orientations until they are in contact.}
\label{contactdis}
\end{center}
\end{figure}

The organization of the remainder of the article is as follows. In
Sec.~\ref{methods}, we describe the methods we employed to measure the
structural relaxation times for the translational and rotational
degrees of freedom for dense liquids composed of anisotropic particles
in 2D.  As a test of our methods, we show in Sec.~\ref{methods} the
results for the collapse of the structural relaxation times for dense
liquids composed of 2D bidisperse disks onto the scaling form
$F_{\pm}(|\phi-\phi_0|,T)$ used in Refs.~\cite{berthier,berthier2} for
3D bidisperse spheres. In Sec.~\ref{results}, we present our novel
results for the structural relaxation times for the rotational and
translational degrees of freedom for dimer- and ellipse-shaped
particles as a function of temperature, packing fraction, and aspect
ratio, and show that the relaxation time data for dimers and
ellipses can be collapsed onto a generalized scaling
function ${\cal F}_{\pm}(|\phi-\phi_0|,T,\alpha)$. In
Sec.~\ref{conclusions}, we summarize our results and identify several
important open questions for future research studies.  The Appendix
includes additional numerical calculations of the system-size
dependence of the structural relaxation times and nematic order as a
function of temperature and packing fraction that supplement the
results in Sec.~\ref{results}.

\section{Methods}
\label{methods}

We performed molecular dynamics (MD) simulations of systems composed
of bidisperse, frictionless rigid dimer- and ellipse-shaped particles
with mass $m$ in 2D that interact via purely repulsive contact forces.
We chose bidisperse systems, where half of the particles are large
with major axis $a_l=1.4$ and half are small with major axis $a_s=1$,
to prevent positional and orientational ordering. We varied the aspect
ratio (ratio of the major and minor axes) $\alpha = a_l/b_l=a_s/b_s$
in the range $1 \leq \alpha \leq 2.2$. (See Fig.~\ref{axes}.) Note
that rigid dimers with $\alpha \gtrsim 2$ possess a small gap between
pairs of monomers that comprise a given dimer. We employed periodic
boundary conditions in a square cell with length $L$ and studied
system sizes in the range $N=64$ to $256$ particles to assess finite-size
effects.

\begin{figure}[htdp]
\begin{center}
\includegraphics[scale=0.4]{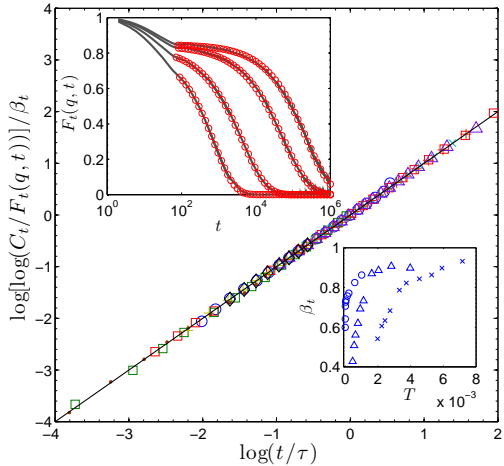}
\caption{Plot of the scaled self-part of the intermediate scattering
function $\log[\log(C_t/F_t(q,t))]/\beta_t$ for the translational degrees
of freedom (symbols) versus $\log (t/\tau_t)$, where $C_t$ is the
prefactor, $\beta_t$ is the stretching exponent, and $\tau_t$ is the
characteristic relaxation time obtained from fits to
Eq.~\ref{stretchedt} for bidisperse disks for packing fractions from
$0.78$ to $0.90$ and temperature from $3\times 10^{-6}$ to $4\times
10^{-3}$.  The solid line has slope $1$. (Inset, top) $F_t(q,t)$
versus $t$ (gray lines) with fits (symbols) to the stretched
exponential form in Eq.~\ref{stretchedt} in the long-time
$\alpha$-decay regime for $\phi=0.83$ and temperatures $T=1\times
10^{-3}$, $9\times 10^{-4}$, $6\times 10^{-4}$, and $5\times 10^{4}$
from left to right. (Inset, bottom) The stretching exponent $\beta_t$
from fits of $F_t(q,t)$ to Eq.~\ref{stretchedt} as a function of
temperature $T$ at $\phi=0.78$ (circles), $0.83$ (triangles), and
$0.90$ (crosses).}
\label{fitcollapse}
\end{center}
\end{figure}

The particles interact via the pairwise, purely repulsive linear
spring potential,
\begin{equation}
\label{pot}
V(r_{ij}) = \frac{\epsilon}{2} \left( 1 - \frac{r_{ij}}{\sigma_{ij}} \right)^2 \Theta\left(1-\frac{r_{ij}}{\sigma_{ij}}\right),
\end{equation} 
where $\epsilon$ is the characteristic energy scale of the repulsive
interaction, $\Theta(x)$ is the Heaviside step function, so that
particles only interact when they overlap, and $\sigma_{ij} $ is the
contact distance between particles $i$ and $j$ that in general depends
on the particle orientations and separation vector ${\vec
r}_{ij}$. For ellipses, ${\vec r}_{ij}$ connects the centers of mass
of ellipses $i$ and $j$ as shown in Fig.~\ref{contactdis}, while for
dimers, ${\vec r}_{ij}$ connects monomers $i$ and $j$ on distinct
dimers $k$ and $l$. For dimers, $\sigma_{ij}=(\sigma_i + \sigma_j)/2$,
where $\sigma_i$ is the diameter of monomer $i$.  The method for
calculating the contact distance between two ellipses is discussed in
detail in Ref.~\cite{schreck_pre}.  Both the large and small ellipses
possess the same moment of inertia $I_e = (a_l/a_s)
mb^2(1+\alpha^2)/16$, while the small dimers have a moment of inertia
$I_{ds} = mb^2(\alpha-1)^2/4$ and large dimers have a moment of
inertia $I_{dl} = mb^2(\alpha-1)^2(a_l/2a_s)^2$.  We chose $\epsilon$
and $b\sqrt{m/\epsilon}$ as the energy and time units.

\begin{figure}[h]
\begin{center}
\includegraphics[scale=0.4]{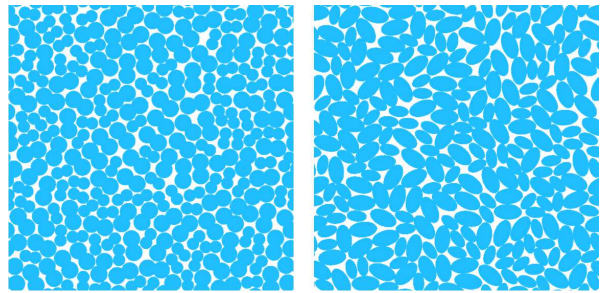}
\caption{(Color online) Snapshots of dense liquid configurations of
bidisperse dimer- (left) and ellipse-shaped particles (right) with
aspect ratio $\alpha=1.8$, packing fraction $\phi=0.82$, and
temperature $T \approx 10^{-4}$.}
\label{snapshots}
\end{center}
\end{figure}

To prepare the systems at a given temperature, we first solve Newton's
equations of motion using the velocity Verlet algorithm with a
velocity rescaling thermostat~\cite{allen} using a time step $\Delta
t=0.01$.  We then equilibrate and study the system dynamics without
the thermostat in the NVE ensemble. To identify the structural
relaxation times, we measure the self-part of the intermediate
scattering function (ISF) for the translational degrees of freedom,
\begin{equation}
F_t(q,t) = \frac{1}{N_l} \left\langle \sum_{j=1}^{N_l} e^{i {\vec q} \cdot [{\vec r}_j(t) - 
{\vec r}_j(0)]} \right \rangle,
\label{position}
\end{equation}
where the wavenumber $q = 2\pi/a_l$, $N_l = N/2$ is the number of
large particles, and $\langle \cdot \rangle$ indicates an average over
the orientations of ${\vec q}$ and time origins.  To quantify relaxation of the rotational
degrees of freedom, we calculated
\begin{equation}
F_r(n,t) = \frac{1}{N_l} \left \langle \sum_{j=1}^{N_l} e^{in
[\theta_j(t) - \theta_j(0)]} \right \rangle,
\label{angle}
\end{equation}
where $\theta_i$ is the angle between the orientation ${\hat u}_i$
of particle $i$ and the horizontal axis and $\langle \cdot \rangle$
indicates an average over time origins.  For most studies, we focused
on $n=1$, in which case relaxation of the ISF occurs when the dimer- or
ellipse-shaped particles flip by $\pi$.

\begin{figure}[h]
\begin{center}
\includegraphics[scale=0.4]{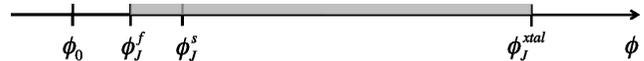}
\caption{Schematic of the possible packing fractions at jamming onset
$\phi_J$ (shaded region) that can be obtained as a function of control
parameters in packing-generation protocols ({\it e.g.} compression or
energy dissipation rates), where $\phi_J^f < \phi_j^s < \phi_J^{\rm
xtal}$.  The most dilute static packings at $\phi_J^f$ are obtained for
infinitely fast quenches, while the densest static packings at $\phi_J^{\rm
xtal}$ (with either complex unit cells or phase-separated
packings for bidisperse systems) are obtained for infinitely slow
quenches.  At finite rates, static packings with intermediate packing
fraction $\phi_J^{s}$ can be obtained.  Note that the current studies
and those in Refs.~\cite{berthier,berthier2} indicate that $\phi_0 <
\phi_J^{f}$.}
\label{slow}
\end{center}
\end{figure}
 
We measured the structural relaxation times for dense liquids composed
of bidisperse disks, dimer-, and ellipse-shaped particles over a wide
range in packing fraction from $0.75$ to $0.94$, temperature from
$10^{-2}$ to $10^{-6}$, and aspect ratio $1 < \alpha < 2$.  In this
regime, the ISF for the
translational and rotational degrees of freedom display stretched
exponential relaxation
\begin{eqnarray}
\label{stretchedt}
F_{t}(q,t) & \approx & C_t e^{-(t/\tau_t)^{\beta_t}} \\
F_{r}(1,t) & \approx & C_r e^{-(t/\tau_r)^{\beta_r}}
\label{stretchedr}
\end{eqnarray}
at long times in the $\alpha$-decay regime, where $C_t$ and $C_r$ are
prefactors, $\beta_t$ and $\beta_r$ are the stretching exponents, and
$\tau_t$ and $\tau_r$ are characteristic relaxation times for the
translational and rotational degrees of freedom, respectively.  In
Fig.~\ref{fitcollapse} we show the ISF for the translational degrees
of freedom for bidisperse disks in 2D with fits to the stretched
exponential form (Eq.~\ref{stretchedt}) at long times.  We find that
the stretching exponent $\beta_t$ decreases with increasing packing
fraction and decreasing temperature (lower inset to
Fig.~\ref{fitcollapse}).  In Sec.~\ref{results}, we will show results
from measurements of the decay of the self-part of the intermediate
scattering function for both the translational and rotational degrees
of freedom for dimer- and ellipse-shaped particles as a function of
$\phi$, $T$, and $\alpha$.  Prior to all measurements, we run MD
simulations of the system for $10$ times the longest relaxation time
$\tau$ to reach metastable equilibrium.

\begin{table}[htdp]
\begin{center}
\begin{tabular*}{0.5\textwidth}{@{\extracolsep{\fill}} c|cccccc}
\hline
	&$\mu$	&$\delta$	&$m(\phi_0)$	&$\phi_0$		&$\phi_J^f$ &$\phi_J^s$\\ \hline
2D  &$1.25\pm0.04$	&$1.9\pm0.1$	&$1.19\pm0.10$	&$0.831\pm0.005$	&$0.838$ & $0.851$\\
3D  &$1.3$			&$2.2\pm0.2$	&$1.43\pm0.13$	&$0.635\pm0.005$	&$0.648$ &$0.662$ \\
\hline
\hline
\end{tabular*}
\end{center}
\caption{Dynamic scaling exponents $\mu$ and $\delta$ obtained from
fitting the structural relaxation time $\tau_t(\phi,T)$ for the
translational degrees of freedom to the form in
Eq.~\ref{scalingformula}, fragility $m(\phi_0)=\mu \delta/2$ at
$\phi_0$, and the packing fractions $\phi_0$, $\phi_J^f$, and
$\phi_J^s$ for dense liquids composed of bidisperse, purely repulsive
disks in 2D and spheres in 3D~\cite{berthier}.}
\label{default}
\end{table}%

We also monitor the positional and orientational ordering of the
systems as a function of decreasing temperature by measuring the bond
orientational order~\cite{steinhardt} and nematic order parameter
\begin{equation}
\label{nematic}
P_2 = \langle 2 \cos^2 (\theta_i - {\overline \theta}) - 1\rangle,
\end{equation}
where ${\overline \theta}$ is the average particle orientation in a
given configuration and the angle brackets indicate an average over
independent configurations.  As shown in the snapshots of the
configurations in Fig.~\ref{snapshots}, we do not find significant
positional or orientational ordering in dense liquids of bidisperse
dimer- and ellipse-shaped particles.

\begin{figure}[h]
\begin{center}
\includegraphics[scale=0.45]{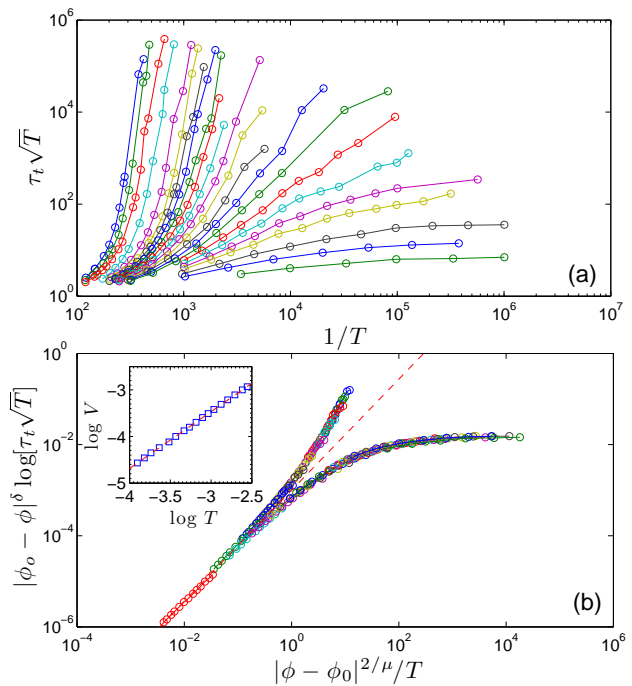}
\caption{(Color online) (a) Structural relaxation time $\tau_t
\sqrt{T}$ versus $1/T$ for dense liquids of bidisperse disks in 2D for
$23$ different packing fractions, $\phi = 0.94$, $0.92$, $0.90$,
$0.88$, $0.86$, $0.85$, $0.845$, $0.84$, $0.835$, $0.83$, $0.825$,
$0.82$, $0.815$, $0.81$, $0.805$, $0.80$, $0.795$, $0.79$, $0.785$,
$0.78$, $0.77$, $0.76$, and $0.75$ from left to right. (b) Collapse of
the data in (a) using Eq. \ref{scalingformula} with $\phi_0 = 0.831,
\mu =1.25$, and $\delta = 1.95$. The dashed line has slope $\mu \delta/2$.
Inset: Scaling relation between the total potential
energy and temperature $V = V_0 T^\mu$ at $\phi=\phi_0\approx 0.83$,
where $V_0 = 2.0$ and $\mu = 1.25$. The dashed line has slope $1.25$.}
\label{2d}
\end{center}
\end{figure}

In previous studies of slow dynamics in dense liquids, Witten
and Berthier~\cite{berthier} identified a dynamical scaling relation,
\begin{equation}
\tau(\phi,T) \sqrt{T} \sim \exp\left[\frac{A}{|\phi_0 - \phi|^\delta}F_\pm\left(\frac{|\phi_0 - \phi|^{2/\mu}}{T}\right)\right],
\label{scalingformula}
\end{equation}
which collapsed the structural relaxation time $\tau(\phi,T)$ from the
self-part of the intermediate scattering function for bidisperse,
purely repulsive spheres in 3D over a wide range of temperature and
packing fraction. The scaling function $F_{\pm}$ assumes two different
forms above and below $\phi_0 \sim 0.635$, and is only a function of
$|\phi_0 -\phi|^{2/\mu}/T$, where $\mu$ is an exponent that controls
the scaling of the potential energy with temperature, $V \approx V_0
T^{\mu}$, near $\phi_0$.  These studies showed that $\tau_t$ displays
hard-sphere dynamics in the zero-temperature limit for $\phi <
\phi_0$, but $\tau_t$ displays super-Arrhenius behavior with
decreasing temperature for $\phi > \phi_0$.  For $\phi \rightarrow
\phi_0$, $x \equiv |\phi_0 - \phi|^{2/\mu}/T \ll 1$ and $F_{\pm}(x)
\sim x^{\delta \mu/2}$ , which implies that $\tau \sim \exp(A/T^{\mu
\delta/2})$ and the fragility $m(\phi_0) = \mu \delta/2$ at $\phi_0$
for $T\rightarrow 0$ is controlled by the scaling exponents $\mu$ and
$\delta$.  The scaling exponent $\delta$ does not depend of the form of
the purely repulsive contact potential. The results for the
scaling exponents $\mu$ and $\delta$, $\phi_0$, and fragility
$m(\phi_0)$ from Ref.~\cite{berthier} are shown in Table~\ref{default}
for dense liquids composed of bidisperse, purely repulsive spheres in
3D.

\begin{figure}
\begin{center}
\scalebox{0.4}{\includegraphics{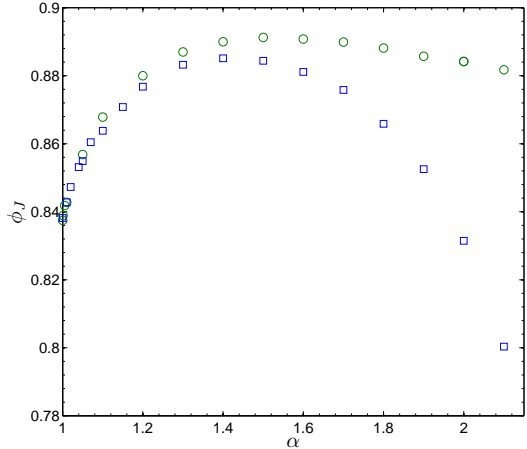}}
\end{center}
\vspace{-0.3in}
\caption{Ensemble averaged packing fraction $\phi_J$ at jamming onset versus
aspect ratio $\alpha$ for $N=480$ bidisperse, frictionless dimer-
(squares) and ellipse-shaped (circles) particles. 
\label{fig1}}
\vspace{-0.1in}
\end{figure}

In Ref.~\cite{berthier}, Berthier and Witten also compared the
critical packing fraction $\phi_0 \sim 0.635$ (which signals the
crossover from hard-sphere dynamics to super-Arrhenius temperature
dependence of $\tau$) to the packing fraction $\phi_J$ at jamming
onset.  Numerous studies have shown that $\phi_J$ for frictionless,
purely repulsive spherical particles depends on control parameters in
the jammed packing-generation protocol ({\it e.g.} the compression and
energy dissipation
rates)~\cite{schreck_order,berthier_order,teitel_order,ciamarra} as
shown schematically in Fig.~\ref{slow}.  In the limit of infinitely
fast rates, the most dilute and disordered mechanically stable
packings at $\phi_J^{f}$ are obtained.  In contrast, in the limit of
infinitely slow rates, the densest crystalline packings are obtained.
At finite rates, intermediate values of the packing fraction at
jamming onset and positional and other types of order can be obtained.
Previous studies have shown for bidisperse systems that disordered
mechanically stable packings at jamming onset can exist over a finite
range of packing fractions ({\it i.e.} from $\phi_J^f$ to $\phi_J^s$
in Fig.~\ref{slow}) with only small and subtle changes in positional
order.  For 3D bidisperse spheres, $\phi_J^f \approx
0.648$~\cite{liu2} and $\phi_J^s \approx 0.662$~\cite{berthier}.  The
scaling analyses in Refs.~\cite{berthier,berthier2} for the structural
relaxation times for 3D bidisperse systems suggest that $\phi_0 < \phi_J^f$.

\begin{figure}[h]
\begin{center}
\includegraphics[scale=0.43]{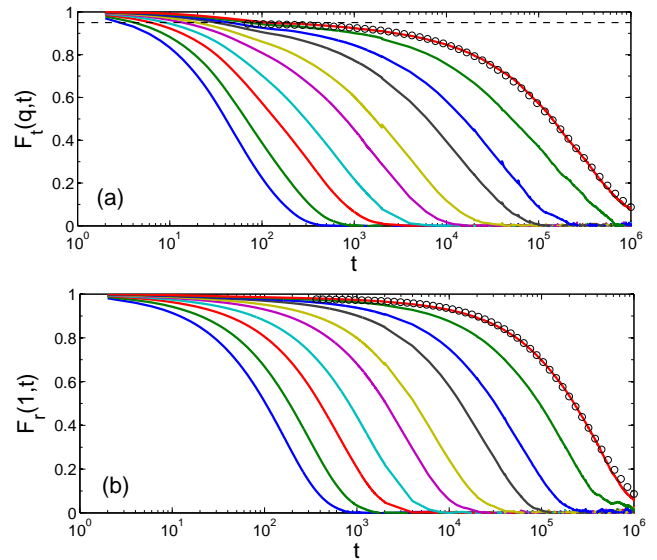}
\caption{(Color online) Self-part of the intermediate scattering
function (ISF) for the (a) translational and (b) rotational ($n=1$)
degrees of freedom for {\it dimers} at aspect ratio $\alpha = 1.6$ and
packing fraction $\phi = 0.86$ over a range of temperatures $T =
5\times10^{-3}$, $3.3\times10^{-3}$, $2.2\times10^{-3}$,
$1.4\times10^{-3}$, $9.3\times10^{-4}$, $6.2\times10^{-4}$,
$4\times10^{-4}$, $2.7\times10^{-4}$, $1.8\times10^{-4}$ and
$1.2\times10^{-4}$ from left to right. The dashed horizontal line in
(a) indicates the approximate value of the plateau in the ISF. The
symbols in (a) and (b) show fits of the ISF to stretched exponential
behavior (Eqs.~\ref{stretchedt} and~\ref{stretchedr}) for long times. }
\label{ISF_dimer}
\end{center}
\end{figure}

Recent experimental~\cite{chaikin,chaikin2} and
computational~\cite{donev,schreck} studies have shown that the average
jammed packing fraction (in the limit of fast quenching rates)
$\phi_J(\alpha)$ for static packings of anisotropic frictionless
particles increases with aspect ratio $\alpha$ from `random close
packing'~\cite{berryman} at $\alpha=1$, reaches a peak near $\alpha_p
\sim 1.5$ (where $\alpha_p$ depends on the specific particle shape and
spatial dimension), and decreases continuously beyond $\alpha_p$.
(See Fig.~\ref{fig1} for a plot of the jammed packing fraction
$\phi_J(\alpha)$ for ellipses and dimers as a function of aspect ratio
in two dimensions (2D).) In these studies, the jammed packings were
created using `fast' packing-generation protocols, and thus these
packings do not possess significant positional or orientational
order~\cite{schreck_prl,schreck_pre,liu}.

\begin{figure}[h]
\begin{center}
\includegraphics[scale=0.44]{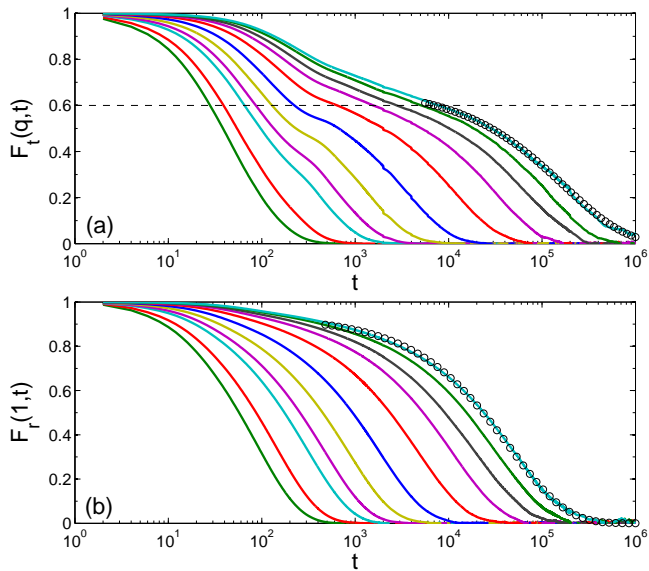}
\caption{(Color online) Self-part of the intermediate scattering
function for the (a) translational and (b) rotational degrees of
freedom ($n=1$) for {\it ellipses} at aspect ratio $\alpha = 1.6$ and
packing fraction $\phi = 0.87$ over a range of temperature $T=
4\times10^{-3}$, $3\times10^{-3}$, $1.8\times10^{-3}$,
$1.4\times10^{-3}$, $1\times10^{-3}$, $6.2\times10^{-4}$,
$3.7\times10^{-4}$, $2.3\times10^{-4}$, $1.7\times10^{-4}$,
$1.2\times10^{-4}$, and $1.0\times10^{-4}$. The dashed horizontal line
in (a) indicates the approximate value of the plateau in the ISF. 
The symbols in (a) and (b) show fits of the ISF to
stretched exponential behavior (Eqs.~\ref{stretchedt}
and~\ref{stretchedr}) for long times.}
\label{ISF_ellipse}
\end{center}
\end{figure}

\begin{figure}[h]
\begin{center}
\includegraphics[scale=0.44]{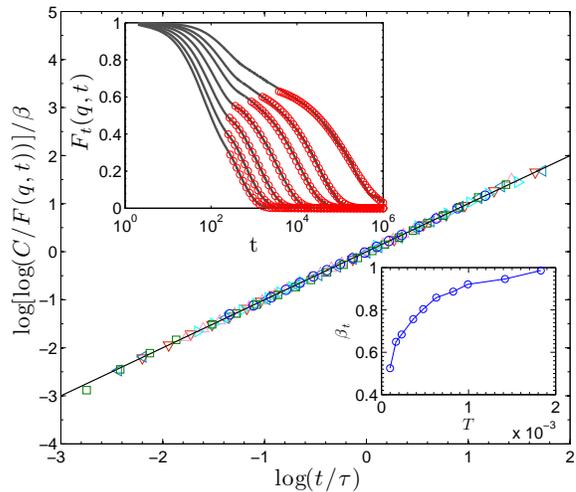}
\caption{Plot of the scaled self-part of the intermediate scattering
function $\log[\log(C/F(q,t))]/\beta$ for the translational and
rotational degrees of freedom (symbols) versus $\log (t/\tau)$, where
$C$ is the prefactor, $\beta$ is the stretching exponent, and $\tau$
is the characteristic relaxation time obtained from fits to
Eqs.~\ref{stretchedt} and~\ref{stretchedr} for ellipses and dimers
at packing fraction $\phi=0.87$ over a range of temperature from $2
\times 10^{-3}$ to $1 \times 10^{-4}$.  The solid line has slope
$1$. (Inset, top) $F_t(q,t)$ versus $t$ (gray lines) with fits
(symbols) to the stretched exponential form in Eq.~\ref{stretchedt} in
the long-time $\alpha$-decay regime for ellipses at $\phi=0.87$ and
temperatures $T=1.8 \times 10^{-3}$, $1. 4\times 10^{-3}$, $1.0 \times
10^{-3}$, $6.2\times 10^{-4}$, $3.7 \times 10^{-4}$, $2.3\times
10^{-4}$, and $1.0 \times 10^{-4}$ from left to right. (Inset, bottom)
The stretching exponent $\beta_t$ from fits of $F_t(q,t)$ to
Eq.~\ref{stretchedt} as a function of temperature $T$ at $\phi=0.87$
(circles) for ellipses.}
\label{collapse2}
\end{center}
\end{figure}

\section{Scaling Analysis for the Structural Relaxation Times}
\label{results}

In this section, we describe our measurements of the structural
relaxation time $\tau$ from the self-part of the intermediate
scattering function for the translational and rotational degrees of
freedom for bidisperse disks, dimer- and ellipse-shaped particles.  We
will investigate the extent to which the scaling relation in
Eq.~\ref{scalingformula} holds for 2D systems as a function of aspect
ratio and determine the variation with aspect ratio of the fragility
at $\phi_0(\alpha)$ that signals the crossover from hard-particle
dynamics to super-Arrhenius temperature dependence. We find no
significant positional or nematic ordering with decreasing temperature
over the full range of packing fraction and aspect ratio. (See
Fig.~\ref{p2} that shows the nematic order parameter $P_2 \lesssim
1/\sqrt{N}$ for $\alpha=1.8$ ellipses over a wide range of packing
fraction and temperature.)

\begin{figure}[h]
\begin{center}
\includegraphics[scale=0.45]{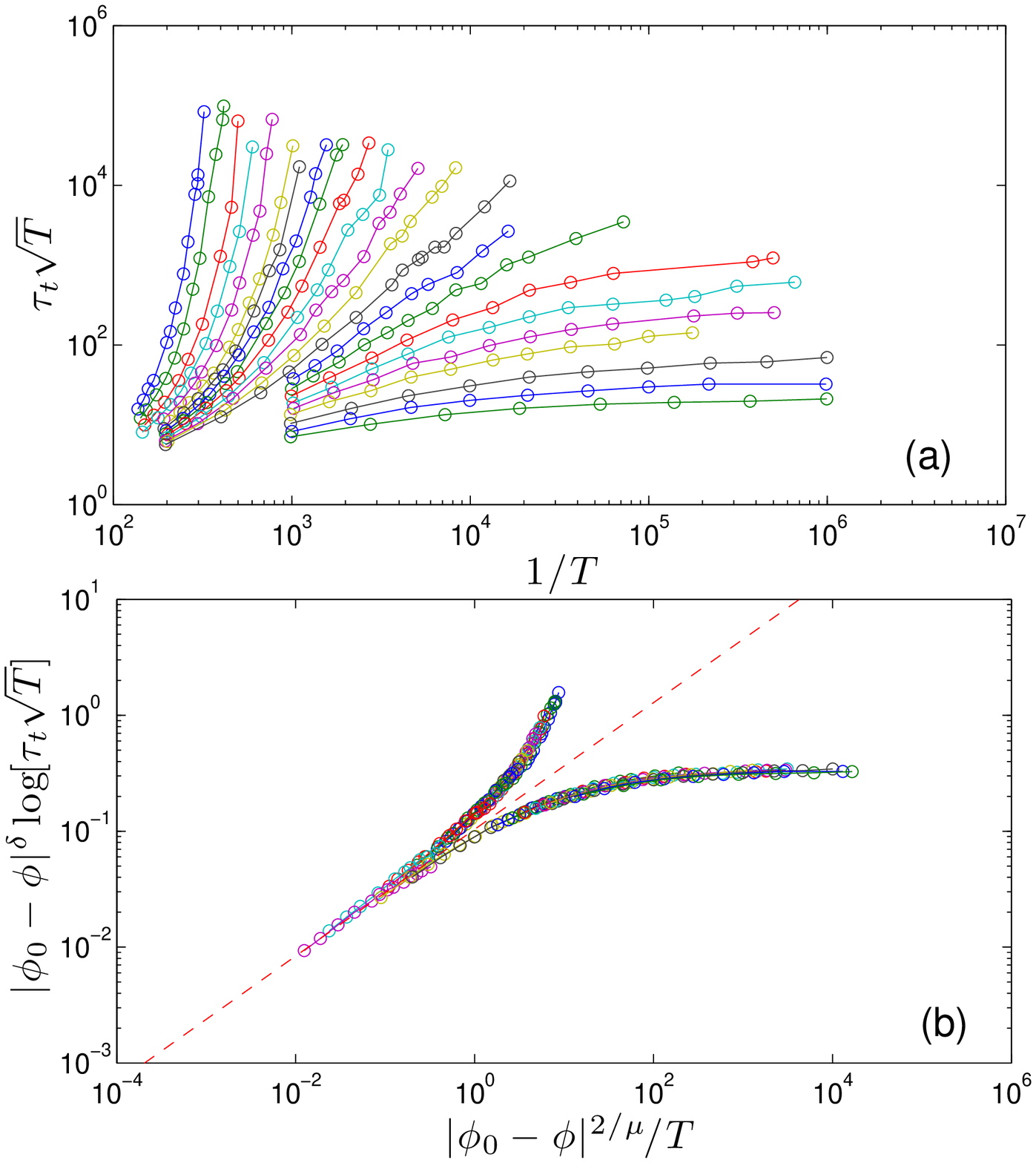}
\caption{(Color online) (a) Structural relaxation time for the
translational degrees of freedom $\tau_t \sqrt{T}$ as a function of
inverse temperature $1/T$ on a log-log scale for dimers at aspect
ratio $\alpha=1.8$ over a range of packing fractions $\phi = 0.95$,
$0.93$, $0.91$, $0.90$, $0.89$, $0.88$, $0.875$, $0.87$, $0.865$,
$0.86$, $0.855$, $0.85$, $0.845$, $0.84$, $0.835$, $0.83$, $0.825$,
$0.82$, $0.815$, $0.81$, $0.80$, $0.79$, and $0.78$ from left to
right.  (b) Collapse of the data in (a) using Eq. \ref{scalingformula}
with $\phi_0 = 0.852, \mu =1.29$, and $\delta = 0.85$. The dashed line
has slope $\mu \delta/2$.}
\label{dimercollapse}
\end{center}
\end{figure}
  
We first focus on the slow dynamics of dense liquids composed of
bidisperse disks in 2D near jamming to compare our results with those
obtained for dense 3D liquids of bidisperse
spheres~\cite{berthier,berthier2}.  The structural relaxation time
$\tau_t$ obtained from the self-part of the intermediate scattering
function is plotted versus the inverse temperature $1/T$ in
Fig.~\ref{2d} (a) over a wide range in packing fraction from $\phi =
0.75$ to $0.94$.  One can clearly identify a change in the form of the
structural relaxation time as the packing fraction increases above
$\phi_0 \approx 0.82-0.84$.  In Fig.~\ref{2d} (b), we collapse the
data in (a) using the scaling form in Eq.~\ref{scalingformula} by
plotting $|\phi_0 - \phi|^{\delta} \log(\tau_t \sqrt{T})$ versus
$|\phi_0 - \phi|^{2/\mu}/T$.  The exponent $\mu$ is obtained from the
power-law scaling of the total potential energy $V$ with temperature
for $\phi \approx \phi_0$~\cite{berthier,berthier2}.  In the inset to
Fig.~\ref{2d} (b), we show that $V \sim T^{\mu}$ over several orders
of magnitude in temperature with an exponent $\mu = 1.25$, which is
similar to the corresponding value for 3D bidisperse spheres,
$\mu_{\rm 3D} = 1.3$~\cite{berthier}.  The scaling form for the
structural relaxation time possesses two distinct branches: $F_+$ for
$\phi > \phi_0$ and $F_-$ for $\phi < \phi_0$. To obtain Fig.~\ref{2d}
(b), we chose the combination of parameters $\phi_0 = 0.831$,
$\mu(\phi_0) = 1.25$, and $\delta = 1.19$ that yielded the tightest
collapse of the $\tau_t$ data in Fig.~\ref{2d} (a). For 2D bidisperse
systems, we find that the fragility at $\phi_0$, $m(\phi_0) = 1.2$,
which is slightly lower than the comparable value, $m(\phi_0) = 1.4$,
for 3D bidisperse spheres. (See Table~\ref{default}.)

\begin{figure}[h]
\begin{center}
\includegraphics[scale=0.44]{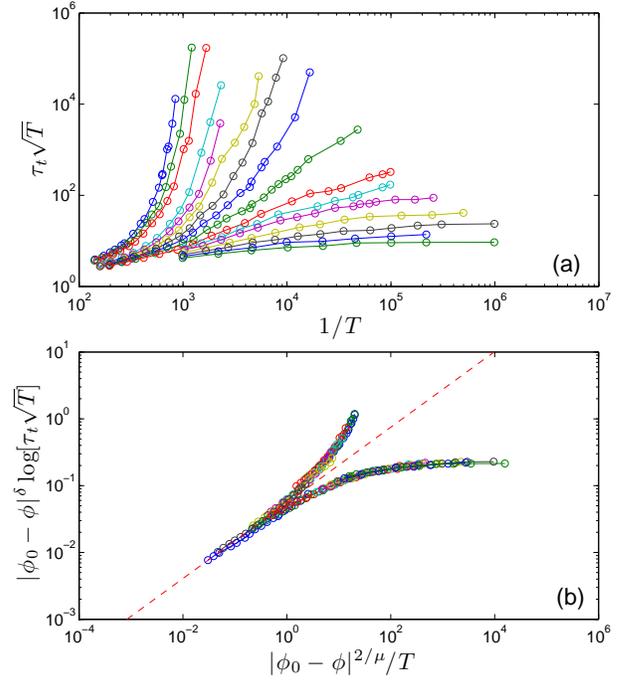}
\caption{(Color online) (a) Structural relaxation time for the
translational degrees of freedom $\tau_t \sqrt{T}$ as a function of
inverse temperature $1/T$ on a log-log scale for ellipses at aspect
ratio $\alpha = 1.8$ over a range of packing fractions $\phi = 0.98$,
$0.96$, $0.94$, $0.92$, $0.91$, $0.90$, $0.89$, $0.88$, $0.87$,
$0.86$, $0.855$, $0.85$, $0.84$, $0.83$, $0.82$, and $0.81$ from left
to right. (b) Collapse of the data in (a) using
Eq. \ref{scalingformula} with $\phi_0 = 0.884, \mu =1.26$, and $\delta
= 0.90$. The dashed line has slope $\mu \delta/2$.}
\label{ellipsecollapse}
\end{center}
\end{figure}

In Figs.~\ref{ISF_dimer} and~\ref{ISF_ellipse}, we show the ISF for
the translational and rotational degrees of freedom for dimers and
ellipses, respectively, over a range of temperatures at comparable
values of the packing fraction $\phi - \phi_J(\alpha)$ for aspect
ratio $\alpha=1.6$ above $\alpha_p$ at which the peak in
$\phi_J(\alpha)$ occurs (Fig.~\ref{fig1}).  For both dimers and
ellipses, the decay of the ISF for the translational and rotational
degrees of freedom at long-times can be described using a stretched
exponential form as shown in Fig~\ref{collapse2}. For $\alpha \gtrsim
\alpha_p$ the rotational degrees of freedom are strongly coupled to
the translational degrees of freedom and $\tau_r \sim \tau_t$. For
both dimers and ellipses, the `plateau' value of the ISF is larger for
the rotational degrees of freedom compared to that for the
translational degrees of freedom.  The only significant difference
between the intermediate scattering functions for dimers and ellipses
in Figs.~\ref{ISF_dimer} and~\ref{ISF_ellipse} is that the plateau
value for the translational degrees of freedom is smaller ($F_t \sim
0.6$ - $0.7$) compared to that for dimers ($F_t \sim 0.95$), which
indicates that the `cage' size for ellipses is larger than that for
dimers.

\begin{figure}[h]
\begin{center}
\includegraphics[scale=0.5]{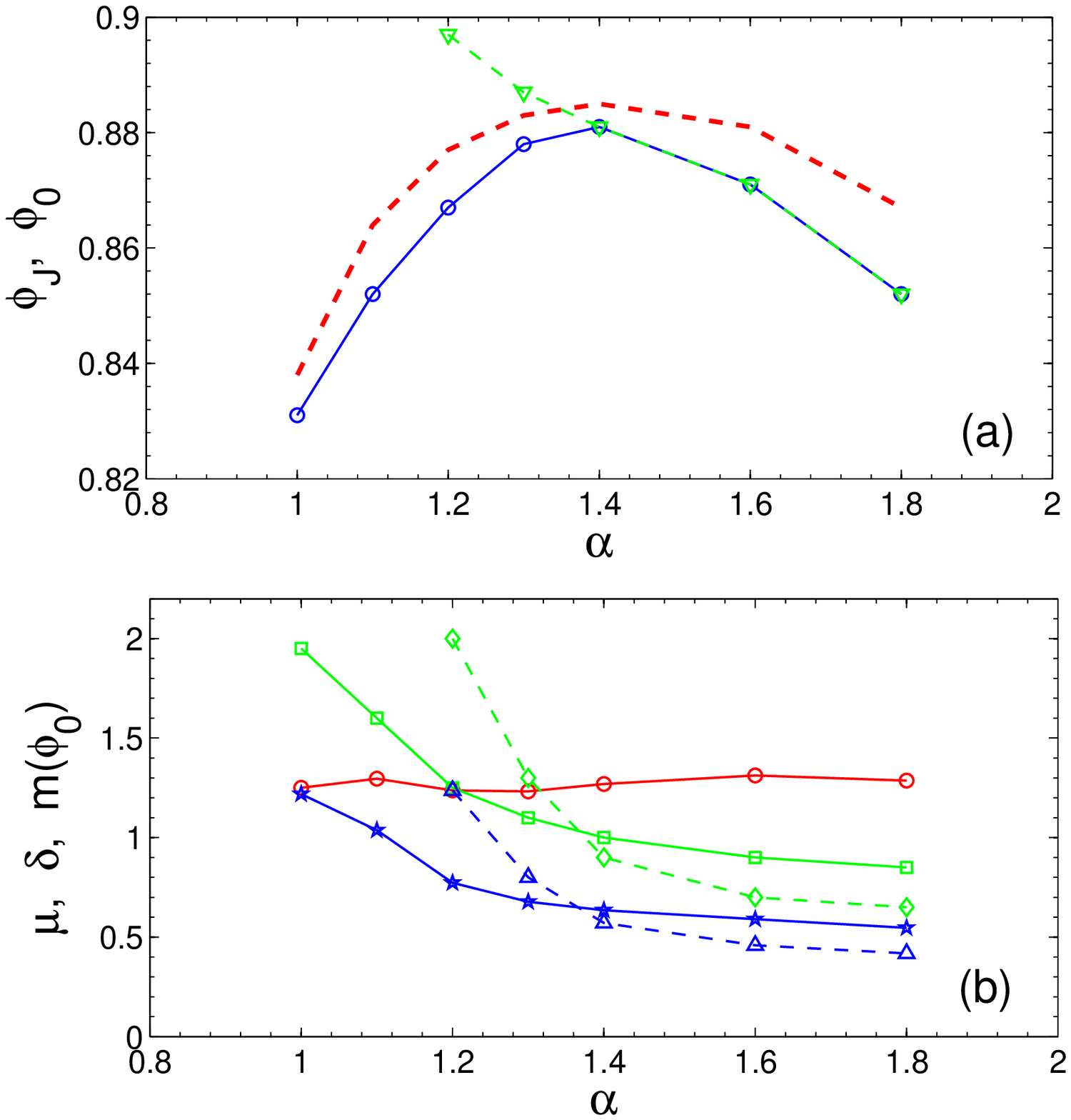}
\caption{(Color online) (a) The characteristic packing fraction
$\phi_0$ that signals the crossover from hard-particle dynamics to
super-Arrhenius temperature dependence for the structural relaxation
times for the translational (circles) and rotational (downward
triangles) degrees of freedom and the packing fraction $\phi_J$ at
jamming onset (dashed line) for dimers as a function of aspect ratio
$\alpha$. (b) The scaling exponents $\mu$ (circles, solid line) and
$\delta$ for translational (squares, solid line) and rotational
(diamonds, dashed line) degrees of freedom, and fragility $m(\phi_0)$
for the translational (asterisks, solid line) and rotational
(triangles, dashed line) degrees of freedom for dimers versus aspect
ratio $\alpha$.}
\label{coldimer}
\end{center}
\end{figure}

\begin{figure}[h]
\begin{center}
\includegraphics[scale=0.5]{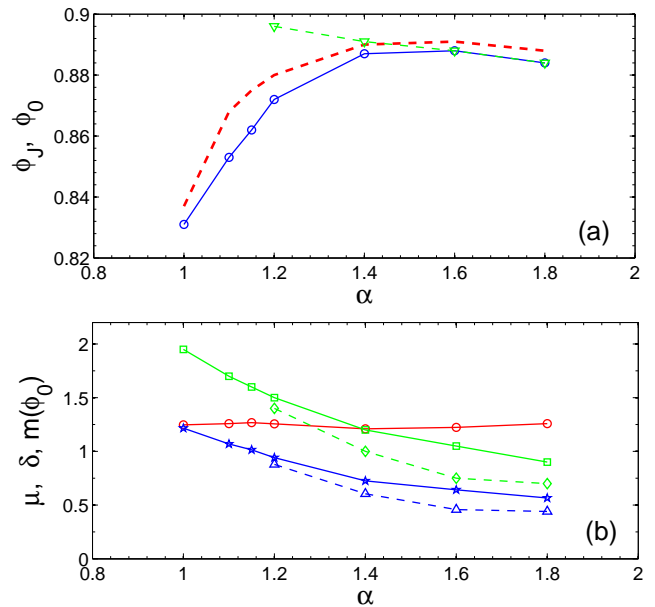}
\caption{(Color online) (a) The characteristic packing fraction
$\phi_0$ that signals the crossover from hard-particle dynamics to
super-Arrhenius temperature dependence for the structural relaxation
times for the translational (circles) and rotational (downward
triangles) degrees of freedom and the packing fraction $\phi_J$ at
jamming onset (dashed line) for ellipses as a function of aspect ratio
$\alpha$. (b) The scaling exponents $\mu$ (circles, solid line) and
$\delta$ for translational (squares, solid line) and rotational
(diamonds, dashed line) degrees of freedom, and fragility $m(\phi_0)$
for the translational (asterisks, solid line) and rotational
(triangles, dashed line) degrees of freedom for ellipses versus aspect
ratio $\alpha$.}
\label{colellipse}
\end{center}
\end{figure}

We use a similar procedure to that employed for bidisperse disks to
collapse the data for the structural relaxation times $\tau_r$ and
$\tau_t$ for dimer- and ellipse-shaped particles using the scaling
form in Eq.~\ref{scalingformula}.  Examples of the collapse of the
structural relaxation times for the translational degrees of freedom
for dimers and ellipses are shown in Figs.~\ref{dimercollapse}
and~\ref{ellipsecollapse}, respectively, at aspect ratio $\alpha =
1.8$. Similar collapse for $\tau_r$ is found for both dimer- and
ellipse-shaped particles.  Also, the quality of the data collapse is
the same for dimers and ellipses.  We plot the characteristic packing
fraction $\phi_0(\alpha)$ that signals the change in scaling form for 
$\tau(\phi,T)$, the scaling exponents
$\mu$ and $\delta$, and the fragility $m(\phi_0)$ for the rotational
and translational degrees of freedom as a function of aspect ratio
$\alpha$ for dimers and ellipses in Figs.~\ref{coldimer}
and~\ref{colellipse}, respectively.

For dimers, the fragility $m(\phi_0)$ for the translational degrees of
freedom decreases by a factor of $\approx 2.5$ with increasing aspect
ratio from $m(\phi_0) \approx 1.2$ at $\alpha = 1$ to $\approx 0.5$ at
$\alpha=1.8$.  The decrease in fragility is caused by the decrease in
the scaling exponent $\delta$ (and thus insensitive to the form of the
purely repulsive contact forces) because $\mu \sim 1.25$ is nearly
constant over this range in aspect ratio.  The fragility for the
rotational degrees of freedom decreases by a similar factor to that
for the translational degrees of freedom for $\alpha \ge 1.2$.  In
addition, in Fig.~\ref{coldimer} (a) we show that the
$\alpha$-dependence of $\phi_0$ for the translational degrees of
freedom matches that for $\phi_J$, {\it i.e.}  nonmonotonic dependence
on $\alpha$ with a peak near $\alpha_p \approx 1.4$, and $\phi_0$
satisfies $\phi_0(\alpha) < \phi_J(\alpha)$.  For $\alpha \ge
\alpha_p$, $\phi_0$ is the same for the translational and rotational
degrees of freedom.  However, for $\alpha < \alpha_p$, $\phi_0$ for
the rotational degrees of freedom increases with decreasing $\alpha$
and becomes larger than $\phi_0$ for the translational degrees of
freedom and $\phi_J$. This strong increase in $\phi_0$ for the
rotational degrees of freedom at small $\alpha$ indicates the
existence of a `rotator', amorphous, solid-like regime~\cite{michele}
at low temperatures, where the translational degrees of freedom are
frozen, but the rotational degrees of freedom are liquid-like.  We
show that these results are nearly independent of system size by
comparing the scaling collapse of the structural relaxation times for
the translational degrees of freedom for systems composed of $N=64$
and $200$ dimers at aspect ratio $\alpha=1.3$ in
Fig.~\ref{system_size}.
  
\begin{figure}[h]
\begin{center}
\includegraphics[scale=0.48]{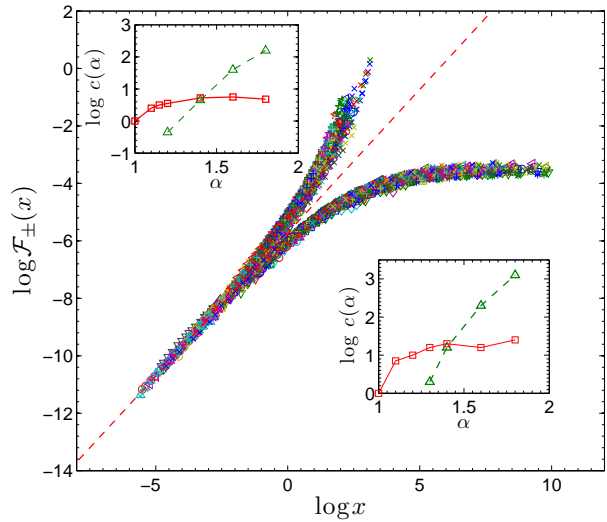}
\caption{The scaling function ${\cal F}_{\pm}(x)$ defined in
Eq.~\ref{rotate} versus $x \equiv |\phi-\phi_0|^{2/\mu}/T$,
that collapses the structural relaxation times for the translational
(Figs.~\ref{dimercollapse} (b) and~\ref{ellipsecollapse} (b)) and
rotational degrees of freedom for ellipses and dimers over a wide
range of aspect ratios. The dashed line has slope $1$.  Insets:
$c(\alpha)$ for the translational (solid) and rotational (dashed)
degrees of freedom for ellipses (top) and dimers (bottom).}
\label{total_collapse}
\end{center}
\end{figure}

We find quantitatively similar behavior for the dependence of the
scaling exponents $\mu$ and $\delta$, fragility $m(\phi_0)$, and
packing fraction $\phi_0$ on aspect ratio for ellipses and dimers as
shown in Fig.~\ref{colellipse}. A minor difference between dimers and
ellipses is the weaker decrease in $\phi_0$ and $\phi_J$ with
increasing aspect ratio $\alpha$.  Again, $\phi_0$ for the rotational
degrees of freedom increases above $\phi_0$ for the translational
degrees of freedom and $\phi_J$ for $\alpha \lesssim \alpha_p$, which
indicates the onset of the `rotator' regime for low aspect ratios.  In
Fig.~\ref{total_collapse} we show that we are able to collapse all of
the structural relaxation time data for the translational and
rotational degrees of freedom for bidisperse dimers and ellipses using
the generalized scaling form
\begin{equation}
{\cal F}_{\pm}(x) = c(\alpha) [F_{\pm}(x)]^{1/m(\phi_0(\alpha))}, 
\label{rotate}
\end{equation}
where $c(\alpha)$ is an $\alpha$-dependent prefactor that is different
for the translational and rotational degrees of freedom and $x\equiv
|\phi-\phi_0|^{2/\mu}/T$.  Thus, we find that the scaling behavior of
the structural relaxation times for dimers and ellipses is nearly
identical over a wide range of aspect ratios, despite the fact that
static packings of dimer-shaped particles are isostatic while static
packings of ellipse-shaped particles are hypostatic at zero
temperature.

\section{Conclusions}
\label{conclusions}

We performed extensive molecular dynamics simulations of the slow
dynamics in dense liquids composed of bidisperse, purely repulsive
dimer- and ellipse-shaped particles in 2D.  We showed that a similar
scaling form to that identified for structural relaxation for
spherical particles~\cite{berthier,berthier2} is able to collapse the
packing fraction and temperature-dependent structural relaxation times
for the rotational and translational degrees of freedom for dimer- and
ellipse-shaped particles over a wide range of aspect ratios $\alpha$
(Fig.~\ref{total_collapse}).  Thus, the dynamical critical point at
$T=0$ and $\phi=\phi_0$ studied in Refs.~\cite{berthier,berthier2} for
dense liquids composed of purely repulsive spheres can be generalized
to $T=0$, $\phi = \phi_0$, and $\alpha=1$, and this dynamical critical
point controls the temperature, packing fraction, and
aspect-ratio-dependent fragility of dense liquids composed of dimer-
and ellipse-shaped particles.  Furthermore, we find qualitatively and
quantitatively similar results for the dynamical scaling exponents,
packing fraction $\phi_0$ that signals the crossover from
hard-particle dynamics to super-Arrhenius temperature dependence, and
fragility $m(\phi_0)$ for dimer- and ellipse-shaped particles.  In
particular, for dimers and ellipses the fragility at $\phi_0$
decreases monotonically with increasing aspect ratio.  Thus, the
microscale differences in shape between dimers and ellipses do not
give rise to important differences in the slow dynamics of structural
relaxation in dense supercooled liquids of anisotropic particles.  In
contrast, static packings of dimers at zero temperature are isostatic,
while static packings of ellipses are hypostatic, which leads to
significant differences in the density of vibrational modes and static
shear modulus at zero temperature~\cite{schreck_prl,schreck_pre}.

These findings suggest several important future studies.  First, we
will investigate possible causes of the decrease in fragility of dense
liquids composed of elongated particles with increasing aspect ratio,
including an increase in the number of inherent
structures~\cite{sastry} and vibrational entropy~\cite{debenedetti}
with increasing $\alpha$.  Second, is there a nonequilibrium,
finite-temperature regime where microscale differences between
anisotropic particle shapes can cause significant differences in the
time-dependent structural and mechanical response of the system?  To
address this question, we will subject jammed packings of dimers and
ellipses to weak thermal fluctuations (much below the temperatures
studied in the present manuscript) and measure the time-dependent
structural and stress relaxation.

\begin{acknowledgments} 
Support from NSF grant numbers DMR-0905880 (BC), DMS-0835742 (TS and
CO), DMR-1006537 (TS and CO), and the Raymond and Beverly Sackler
Institute for Biological, Physical, and Engineering Sciences (TS) is
acknowledged.  This work also benefited from the facilities and staff
of the Yale University Faculty of Arts and Sciences High Performance
Computing Center and NSF Grant No. CNS-0821132 that partially funded
acquisition of the computational facilities. BC also acknowledges the
hospitality of the Schlumberger-Doll Research Center where part of
this work was performed.
\end{acknowledgments}

\appendix 

\section{Supplementary Results}
\label{rotation}

\begin{figure}[htdp]
\begin{center}
\includegraphics[scale=0.4]{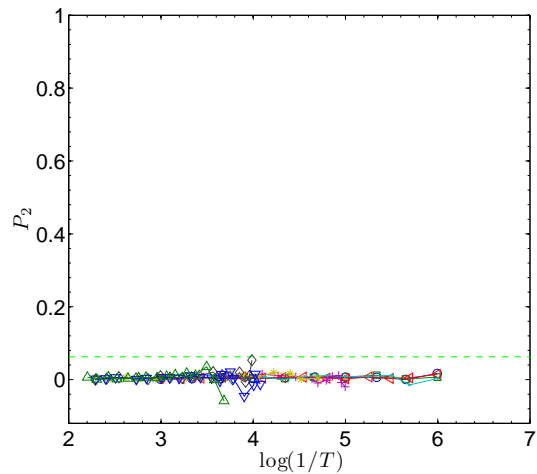}
\caption{Nematic order parameter $P_2$ versus $\log (1/T)$ for dense liquids
composed of $N=200$ ellipses at aspect ratio $\alpha = 1.8$ averaged over
$100$ configurations for a range of packing fractions $\phi=0.81$
(open circles), $0.82$ (squares), $0.83$ (leftward triangles), $0.84$
(filled circles), $0.85$ (pluses), $0.86$ (asterisks), $0.87$
(diamonds), $0.88$ (downward triangles), and $0.90$ (upward
triangles).  The dashed horizontal line indicates $P_2 = 1/\sqrt{N}$.}
\label{p2}
\end{center}
\end{figure}

In this section, we provide additional numerical calculations that
supplement the results presented in Sec.~\ref{results}.  In
Fig.~\ref{p2}, we show the average nematic order $P_2$ (Eq.~\ref{nematic})
for dense liquids composed of ellipses with $\alpha = 1.8$ versus
$1/T$ over a wide range of packing fractions above and below $\phi_0$.
$P_2 \lesssim 1/\sqrt{N}$ for all systems studied, which indicates that
the ellipse orientations are disordered.  We obtain similar results to
those in Fig.~\ref{p2} for dense liquids of bidisperse dimers over the range 
of $T$, $\phi$, and $\alpha$ considered. 

\begin{figure}[htdp]
\begin{center}
\includegraphics[scale=0.4]{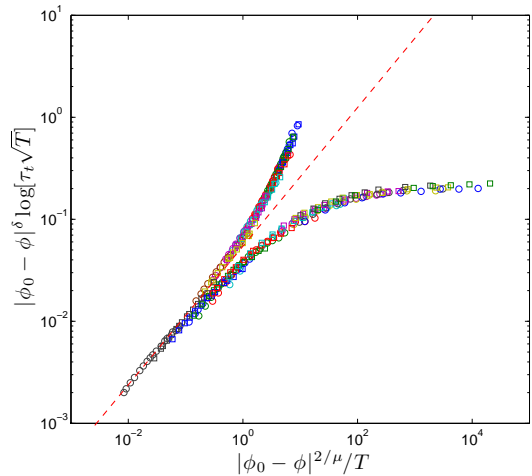}
\caption{Scaled structural relaxation times for the translational
degrees of freedom $|\phi_0 - \phi| \log \tau_t \sqrt{T}$ versus the
scaled temperature $|\phi_0 - \phi|^{2/\mu}/T$ on a log-log scale for
dimers at $\alpha=1.3$ for $N=64$ (squares) and $200$ (circles)
particles. The system-size dependence of the scaling parameters is
weak: $\phi_0=0.878$ ($0.876$); $\mu=1.23$ ($1.25$); and $\delta =
1.1$ ($1.1$) for $N=256$ ($64$) particles. The dashed line has slope
$\mu \delta/2$ for $N=256$.}
\label{system_size}
\end{center}
\end{figure}

In Fig.~\ref{system_size}, we investigate the effect of system size on
the structural relaxation times.  We show that the scaled structural
relaxation times for the translational degrees of freedom for dense
liquids composed of dimers for two system sizes ($N=64$ and $200$) are
nearly identical.  Similar results to those in Fig.~\ref{system_size}
for dimers hold over the full range of aspect ratio considered.  We
also find that there are negligible system-size effects for the
structural relaxation times for ellipses with $N \ge 64$ for $1 < \alpha < 2$.

\begin{references}

\bibitem{debenedetti}
P. G. Debenedetti and F. H. Stillinger, {\it Nature} {\bf 410} (2001) 259. 

\bibitem{angell}
C. A. Angell, {\it Science} {\bf 267} (1995) 1924. 

\bibitem{haxton}
M. Schmiedeberg, T. K. Haxton, S. R. Nagel, and A. J. Liu, {\it Europhys. 
Lett.} {\bf 96} (2011) 36010. 

\bibitem{haxton2}
T. K. Haxton, M. Schmiedeberg, and A. J. Liu, {\it Phys. Rev. E} 
{\bf 83} (2011) 031503. 

\bibitem{berthier}
L. Berthier and T. A. Witten, {\it Europhys. Lett.} {\bf 86} (2009) 10001.

\bibitem{berthier2}
L. Berthier and T. A. Witten, {\it Phys. Rev. E} {\bf 80} (2009) 021502.  

\bibitem{liu2}
C. S. O'Hern, L. E. Silbert, A. J. Liu, and S. R. Nagel, {\it Phys. Rev. E}
{\bf 68} (2003) 011306. 

\bibitem{schreck_order}
C. F. Schreck, C. S. O'Hern, and L. E. Silbert, {\it Phys. Rev. E} 
{\bf 84} (2011) 011305. 

\bibitem{letz} M. Letz, R. Schilling, and A. Latz, {\it Phys. Rev. E}
  {\bf 62} (2000) 5173.

\bibitem{pfleiderer} P. Pfleiderer, K. Kilinkovic, and T. Schilling,
  {\it Europhys. Lett.} {\bf 84} (2008) 16003.

\bibitem{zhang1} R. Zhang and K. S. Schweizer, {\it J. Chem. Phys.} {\bf 133}
(2010) 104902.

\bibitem{kammerer} S. Kammerer, W. Kob, and R. Schilling, {\it Phys.
  Rev. E} {\bf 56} (1997) 5450.

\bibitem{kammerer1} S. Kammerer, W. Kob, and R. Schilling, {\it
  Phys. Rev. E} {\bf 58} (1998) 2131.

\bibitem{kammerer2} S. Kammerer, W. Kob, and R. Schilling, {\it
  Phys. Rev. E} {\bf 58} (1998) 2141.

\bibitem{schreck} C. F. Schreck, N. Xu, and C. S. O'Hern, 
{\it Soft Matter} {\bf 6} (2010) 2960.

\bibitem{witten}
A. V. Tkachenko and T. A. Witten, {\it Phys. Rev. E} {\bf 60}
(1999) 687.

\bibitem{schreck_pre} C. F. Schreck, M. Mailman, B. Chakraborty, and C. S. 
O'Hern, {\it Phys. Rev. E} {\bf 85} (2012) 061305.

\bibitem{allen} M. P. Allen and D. J. Tildesley, {\it Computer
Simulations of Liquids} (Oxford University Press, New York, 1987).

\bibitem{steinhardt}
P. J. Steinhardt, D. R. Nelson, and M. Ronchetti, {\it Phys. Rev. B} 
{\bf 28} (1983) 784. 

\bibitem{berthier_order}
P. Chaudhuri, L. Berthier, and S. Sastry, {\it Phys. Rev. Lett.} {\bf 104}
(2010) 165701. 

\bibitem{teitel_order}
D. V{\aa}gberg, P. Olsson, and S. Teitel, {\it Phys. Rev. E} {\bf 83}
(2011) 031307. 
 
\bibitem{ciamarra}
M. P. Ciamarra, M. Nicodemi, and A. Coniglio, {\it Soft Matter} {\bf 6} (2010)
2871. 

\bibitem{chaikin}
W. N. Man, A. Donev, F. H. Stillinger, M. T. Sullivan, W. B. Russel, D. 
Heeger, S. Inati, S. Torquato, and P. M. Chaikin, {\it Phys. Rev. Lett.}
{\bf 94} (2005) 198001.

\bibitem{chaikin2}
A. Donev, I. Cisse, D. Sachs, E. A. Variano, F. H. Stillinger, R. Connelly, 
S. Torquato, and P. M. Chaikin, {\it Science} {\bf 303} (2004) 990.

\bibitem{donev}
A. Donev, R. Connelly, F. H. Stillinger and S. Torquato,
{\it Phys. Rev. E} {\bf 75} (2007) 051304.

\bibitem{berryman}
J. G. Berryman, {\it Phys. Rev. A} {\bf 27} (1983) 1053. 

\bibitem{schreck_prl} M. Mailman, C. F. Schreck, C. S. O'Hern, and
B. Chakraborty, {\it Phys. Rev. Lett.} {\bf 102} (2009) 255501.

\bibitem{liu} Z. Zeravcic, N. Xu, A. J. Liu, S. R. Nagel, and W. an Saarloos, 
{\it Europhys. Lett.} {\bf 87} (2009) 26001.  

\bibitem{michele}
C. De Michele, R. Schilling, and F. Sciortino, {\it Phys. Rev. Lett.} {\bf 98}
(2007) 265702. 

\bibitem{sastry}
S. Sastry, {\it Nature} {\bf 409} (2000) 164. 
  
\end{references}
\end{document}